\documentclass[aps,showpacs,preprintnumbers,amsmath,amssymb,nofootinbib]{revtex4}

\usepackage{graphicx}

\def\laq{~\raise 0.4ex\hbox{$<$}\kern -0.8em\lower 0.62
ex\hbox{$\sim$}~}
\def\gaq{~\raise 0.4ex\hbox{$>$}\kern -0.7em\lower 0.62
ex\hbox{$\sim$}~}

\def\be{\begin{equation}}
\def\ee{\end{equation}}
\def\bea{\begin{eqnarray}}
\def\eea{\end{eqnarray}}
\newcommand{\nn}{\nonumber}

\def \ra {\rightarrow}
\def \l {\lambda}
\def \L {\Lambda}
\def \d {\delta}
\def \D {\Delta}

\def \G {\Gamma}
\def \g {\gamma}
\def \s {\sigma}

\def \e {\epsilon}

\def \Om {\Omega}
\def \k {\kappa}

\def \rDM {r_{\rm DM}}
\def \rhoDM {\rho_{\rm DM}}
\def \uDM {u_{\rm DM}}
\def \rS {r_{\rm S}}

\begin{document}

BA-TH 654-12

\title{Solar System constraints on local dark matter density}

\author{Giuseppe De Risi}
\email{giuseppe.derisi@port.ac.uk}
\affiliation{Dipartimento di Fisica, Universit\`a degli studi di Bari ``Aldo Moro'', \\
Via G. Amendola 173, 70126 Bari, Italy}

\author{Tiberiu Harko}
\email{harko@hkucc.hku.hk} \affiliation{Department of Physics and
Center for Theoretical and Computational Physics, The University
of Hong Kong, Pok Fu Lam Road, Hong Kong}

\author{Francisco S.~N.~Lobo}
\email{flobo@cii.fc.ul.pt} \affiliation{Centro de
Astronomia e Astrof\'{\i}sica da
             Universidade de Lisboa, Campo Grande, Ed. C8 1749-016 Lisboa,
             Portugal}

\date{\today}

\begin{abstract}

We study how the classical tests of general relativity are
modified by the presence of a subdominant dark matter halo in the
solar system. We use a general formalism to calculate the
corrected expression for the relevant parameters, and obtain bound
plots for the mean energy density and the dimension of the dark
matter halo. Our results seem to favor a density profile peaked at
the center of the solar system.

\end{abstract}


\maketitle

\section{Introduction}
\label{Introduction}

It is widely accepted that more than $80\%$ of the matter of our
universe is ``dark'', a term indicating either that it does not
interact with photons, or our lack of knowledge of the profound
nature of its constituents. Since the seminal observations by Oort \cite{Oort}
and Zwicky \cite{Zwicky} of the early XX century, there have been several
independent evidences that  some sort of non-baryonic
dark matter may exist: the observations of the galactic rotation curves
\cite{Rubin:1970zza} and strong lensing data from Hubble space
telescope \cite{Hudson:1997bj} can be accounted for by the single
hypothesis that galaxies are surrounded by a much larger dark
matter halo, which contains most of the mass of the galaxies. This is
remarkably consistent with the $\L$CDM model of our universe,
suggested by the WMAP observations \cite{Komatsu:2010fb}, which
assumes that the mass energy density in our universe is about
$30\%$ of the critical energy density, while the observable matter
is only $4 \%$, and with various observations on clusters  of galaxies, which
suggest a ``mass-to-light'' ratio much higher than $1$. In
addition, successful theoretical models and numerical simulations
of the formation of structures requires the presence of dark
matter.

However, if the existence of dark matter is (almost) commonly
accepted, its distribution poses serious challenges. In fact,
theoretical models and numerical simulations predict, both for
clusters of galaxies \cite{Navarro:1996gj} and for single galaxies
themselves \cite{Salucci:2002jg,Reed:2003hp} that the radial
density profile is quite steep and peaked at the center of the
galaxy. But other evidences suggest, on the contrary, that the
density profile should be flat, or even shallow
\cite{de_Blok:2001fe}. For a comprehensive review of the density
profile problem (and many others issue related to dark matter),
see \cite{Bertone:2004pz}.

The possibility that the galactic dynamics of massive test
particles may be understood without the need for dark matter was
also considered in the framework of $f(R)$ modified theories of
gravity \cite{Boehmer:2007kx,geomdm,Bohmer:2007fh}. In particular,
the vacuum gravitational field equations in $f(R)$ gravity, in the
constant velocity region, and the general form of the metric
tensor is derived in a closed form was analyzed
\cite{Boehmer:2007kx}. The resulting modification of the
Einstein-Hilbert Lagrangian is of the form $R^{1+n}$, with the
parameter $n$ expressed in terms of the tangential velocity.
Therefore, it was concluded that in order to explain the motion of
test particles around galaxies only requires very mild
deviations from classical general relativity, and that modified
gravity can explain the galactic dynamics without the need of
introducing dark matter. In the context of modified gravity, the
virial theorem in $f(R)$ gravity was generalized by using the
collisionless Boltzmann equation \cite{Bohmer:2007fh}. It was
found that supplementary geometric terms in the modified Einstein
equation provides an effective contribution to the gravitational
energy. Furthermore, the total virial mass was found to be
proportional to the effective mass associated with the new
geometrical term, which may account for the well-known virial
theorem mass discrepancy in clusters of galaxies. The model
considered in \cite{Bohmer:2007fh} predicts that the geometric
mass and its effects extend beyond the virial radius of the
clusters. Thus, it was shown that the $f(R)$ virial theorem can be
an efficient tool in observationally testing the viability of this
class of generalized gravity models.

It is interesting to note that recently the possibility that dark
matter is a mixture of two non-interacting perfect fluids, with
different four-velocities and thermodynamic parameters was
extensively analyzed \cite{Harko:2011nu}.
By assuming a non-relativistic kinetic model for the dark matter
particles, the density profile and the tangential velocity of the
dark matter mixture were obtained by numerically integrating the
gravitational field equations. The cosmological implications of
the model were also briefly considered, and it was shown that the
anisotropic two-fluid model isotropizes in the large time limit.
In fact, this model was further explored in \cite{Harko:2011kw},
by assuming that the two dark matter components are pressureless,
non-comoving fluids. For this particular choice of the equations
of state the dark matter distribution can be described as a single
anisotropic fluid, with vanishing tangential pressure, and
non-zero radial pressure.  The
general, radial coordinate dependent, functional relationship
between the energy density and the radial pressure was also
determined, and it was shown to differ from a simple barotropic
equation of state.

Presently, it is considered that the upper bound of the dark matter density in the Solar System is $\rho _{DM}^{SS}\approx 2\times 10^{-19}$ g/cm$^3$$=10^5 \;\left({\rm GeV}/c^2\right)$ cm$^{-3}$ \cite{Iorio}. The presence of a local dark matter halo could have some influence on the motion
of the objects of the solar system \cite{Sereno:2006mw,Iorio:2006cn,Frere:2007pi}. The problem of the density and distribution of the dark matter in the Solar System is of fundamental importance, not only from a pure theoretical point of view, but also for the design of the dark matter particle detectors. Recently an analysis of the kinematics of 412 stars at 1¡V4 kpc from the Galactic midplane performed in \cite{Bidin} has  derived a local density of dark
matter that is an order of magnitude below standard expectations, of the order of $10^{-3}M_{\odot}$ pc$^{-3}$, or 0.04 GeV/cm$^3$. This result was contested in \cite{Trem}, where it was claimed that it arises from the invalid assumption that the mean azimuthal velocity of the stellar tracers is independent of the Galactocentric radius at all heights.  The assumption of constant mean azimuthal velocity is physically implausible, since it requires the circular velocity to drop more steeply than allowed by any plausible mass model, with or without dark matter, at large heights above the mid-plane \cite{Trem}. Using the correct approximation that the circular velocity curve is flat in the mid-plane, it was found  that the data imply a local dark-matter density of $0.3\pm 0.1$ Gev/cm$^3$  \cite{Trem}, consistent with the standard estimates. A new method for the determination of the local disk matter and dark halo matter density was proposed in \cite{Garbari}.  The method assumes only that the disc is locally in dynamical equilibrium, and that the 'tilt' term in the Jeans equations is small up to 1 kpc above the plane. By using this approach a local dark matter density of $\rho _{DM} = 0.025^{+0.014}_{-0.013}$ $M_{\odot}$/pc$^3$ ($0.95^{+0.53}_{-0.49}$ GeV/cm$^3$)
at 90\% confidence level, assuming no correction for the non-flatness of the local rotation curve, and $\rho _{DM} = 0.022^{+0.015}_{-0.013}$ $M_{\odot}$/pc$^3$ ($0.85^{+0.57}_{-0.50}$ GeV/cm$^3$), if the correction is included. The obtained lower bound for the local dark matter density is larger than the standard adopted value, and it is inconsistent with the data obtained by  extrapolating rotation curves that
assume a spherical halo.

In this paper we will consider a different approach to the problem of the dark matter density in the Solar System. If
the dark matter density profile follows (roughly) the baryonic matter
distribution (for example, because of gravitational interaction), it is
plausible to assume that there can be specific features at a much lower
scale with respect to the galactic one. In particular, due to some accretion processes, there can be an excess of dark matter around compact objects such as the Sun. Numerical simulation, of
course, do not have the resolution to test this hypothesis, but
some speculations have been made. A brief review of this
particular issue is given in \cite{Adler:2009ir}.  Our aim is to
study how classical tests of general relativity, i.e., the
precession of the perihelion of the planet Mercury,  the deflection of light rays
and the delay of radio signals passing close to the Sun, are
modified by the presence of a local, subdominant dark matter
energy density distribution. The Solar System tests are very powerful tools for testing different gravitational models, as well as the theoretical extensions of General Relativity, and they have been applied
recently to a variety of contexts \cite{Boehmer:2008zh,Harko:2009qr,arXiv:0910.3800}.

The present paper is organized as follows.  In Section~\ref{metric}, we derive an
approximate expression for the space-time metric of the Solar
System in the presence of dark matter. In Section
\ref{DM_tests} we use this expression, and the formalism of
Section~\ref{general} to evaluate the corrections due to the presence of dark matter to the relevant
parameters in the Solar System gravitational tests. Finally in Section
\ref{Comments}, we discuss our results and draw our conclusions. In the Appendix we
describe the general formalism used to evaluate the
observational parameters of physical interest.

\section{Metric of a static dark matter halo surrounding the Sun}
\label{metric}

Throughout this work, we assume a general metric of the form
\be ds^2 = A(r) c^2dt^2 -B(r) dr^2 -r^2 d\Om^2 \,, \label{gen_metric}
\ee
which will be used to obtain the general expressions for the
classical tests of general relativity, namely, for the precession
of the perihelion of Mercury, for the deflection of light rays
passing close to the Sun and for the radar echo delay observations. For
the Schwarzschild metric, giving the spherically symmetric vacuum solution of the Einstein gravitational field equations,
 the functions $A(r)=A_0(r)$ and $B(r)=B_0(r)$ are given by
\be A_0(r) = B_0(r)^{-1} = 1 - \frac{2GM}{c^2r} = 1 - \frac{\rS}{r}
\label{schw_metric}
\ee
where $\rS = 2GM/c^2$ is the Schwarzschild radius, with $M$ the mass of the central object.

We begin this Section by deriving an approximate expression
for the metric of a space-time filled by {\it static} dark matter
around a star, by using an ansatz of the form (\ref{gen_metric}).
We assume that the dark matter source is described by a perfect fluid
stress-energy tensor
\be T_{\mu \nu} = \left[ \rho(r) + \frac{p(r)}{c^2}\right]u_\mu
u_\nu - p(r) g_{\mu \nu}\,, \label{DM_set}
\ee
where $\rho$ and $p$ are related by the equation of state $p = \g
\rho c^2$, $\gamma =$ constant (later on, we will set $\g = 0$ according to the general
equation of state of the matter source). The Einstein field equations
reduce to two independent differential equations (we set $8\pi
G/c^4 = \k)$
\bea
\frac{d}{dr}\left[ r\left( 1- \frac{1}{B(r)}\right)\right] &=&
\k c^2 r^2 \rho(r), \label{DM_eq1}  \\
\frac{-r A'(r) + A(r)\left[B(r) - 1\right]}{r^2 A(r) B(r)} &=& -\k p(r).
\label{DM_eq2}
\eea

We assume that energy density and pressure of dark matter are very
small, $\rho \equiv \rho_1$ and $p \equiv p_1$, and hence the presence
of dark matter can be regarded as a small perturbation of the
vacuum state. Therefore we can expand the metric around the
Schwarzschild solution as
\bea
A(r) = A_0(r) + A_1(r),&~~~~~&A_1(r) \ll A_0(r)\,, \\
B(r) = B_0(r) + B_1(r),&~~~~~&B_1(r) \ll B_0(r)\,,
\label{metric_exp}
\eea
with $A_0$ and $B_0$ given by Eq. (\ref{schw_metric}). Then, after
some algebraic manipulation, using the leading order equation of
motion  and neglecting higher order terms, Eq.(\ref{DM_eq1})
becomes
\be
\frac{d}{dr}\left[ r \frac{B_1(r)}{B_0^2(r)}\right] = \k c^2
r^2 \rho_1(r). \label{DM_eq_approx1}
\ee
By defining the effective mass function (see, for example,
\cite{Hobson:2006se})
\be m(r) = 4\pi \int_0^r \bar{r}^2 \rho(\bar{r}^2) d\bar{r}\,,
\label{m} \ee
from Eq.~(\ref{DM_eq_approx1}) we obtain the function $B_1(r)$ as
\be B_1(r) = B_0^2(r) \frac{\rS}{r} \frac{m_1(r)}{r}, \label{B_1}
\ee
where we have
expressed the gravitational coupling constant in terms of the mass
and of the Schwarzschild radius.
By using the definition of the effective mass and the equation of
state of the matter, we can relate $\rho$ and $p$ with the derivative of $m(r)$.
Then, after some algebra and using again leading order equations
of motion and solutions, Eq.~(\ref{DM_eq2}) becomes
\be A_1'(r) -
\frac{\frac{\rS}{r}}{1-\frac{\rS}{r}}\frac{A_1(r)}{r} =
\frac{1}{M}\frac{\rS}{r}\left[ \frac{\g}{c^2}m'_1(r) +
\frac{1}{1-\frac{\rS}{r}} \frac{m_1(r)}{r} \right],
\label{DM_eq_approx2} \ee
where the prime denotes a derivative with respect to $r$.

Now, using the conservation equation to get another independent
equation relating the metric and the effective mass would be an
unnecessary effort, since the barotropic assumption is too
constrictive to get a realistic energy distribution, and the
resulting equations of motion are too difficult to handle. We will
instead use, for the dark matter energy density, an effective
distribution, inspired by the Navarro-Frenk-White profile for the
energy density of the galactic halos \cite{Frere:2007pi}
\be \rho_1(r) \equiv \rhoDM \left( \frac{r}{\rDM}\right)^{-\l} \,,
\label{DM_profile} \ee
where $\rhoDM$ and $\rDM$ are constant arbitrary parameters
related to the mean density and the dimension of the star halo, and $\lambda $ is assumed to be a constant. We
set $\g = 0$, as usual for pressureless matter sources. Then the effective mass
function and the $B_1$ metric function can be easily evaluated,
and are provided by the following relationships
\bea
m(r) &=& \frac{4\pi}{3-\lambda} \rDM^3 \rhoDM \left( \frac{r}{\rDM}
\right)^{3-\l}, \label{m_sol} \\
B_1(r) &=& \frac{4\pi}{3-\lambda} \frac{1}{(1 - \frac{\rS}{r})^2}
\frac{\rS}{\rDM} \frac{\rDM^3 \rhoDM}{M} \left( \frac{r}{\rDM}
\right)^{2-\l}. \label{B_1_sol} \eea
A closed expression for $A_1$ can not be found. However, the
general solution of Eq. (\ref{DM_eq_approx2}) is given by
\be A_1(r) = \frac{4 \pi}{3- \l}\frac{\rS}{\rDM} \frac{\rDM^3
\rhoDM}{M} \left( 1 - \frac{\rS}{r} \right) \int \frac{1}{\rDM}
\frac{\left(r/\rDM \right)^{1-\l}}{\left( 1 -
\rS/r \right)^2}dr\,. \label{A_1_sol} \ee

Though it is reasonable to assume $\l > 0$, there is no need in
principle to set another limit to $\l$, since our approximation
holds only outside the Sun, and a matter distribution like
(\ref{DM_profile}) can be trusted up to a cutoff radius (for
example the virial radius in the NFW distribution). Nevertheless,
the analysis we are proposing makes sense only if the contribution
to the metric around the Sun from dark matter is actually
localized in the Solar System. Thus, we do expect the corrections
to the metric functions to vanish as $r \ra +\infty$, and so we
will assume for simplicity $\l > 2$.

In the next Section, we will use the approximated metric we have
deduced to see how the classical tests of general relativity
constrain the form of the dark matter distribution.

\section{Classical tests of General Relativity in the Dark Matter halo}
\label{DM_tests}

In Appendix \ref{general}, we briefly review the formalism used in
studying the classical tests of general relativity, which we have
included for self-consistency and self-completeness. The
formalism is completely general, and can be used to treat any
static spherically symmetric metric \cite{Harko:2009qr,arXiv:0910.3800}.

In this Section, we will use the metric given by Eqs.~(\ref{B_1_sol}) - (\ref{A_1_sol}) to study how the results of the classical tests of General Relativity (perihelion precession,
light deflection and radar echo delay) are modified in the Solar System due to the presence of the dark matter. Let us first write down the metric functions in terms of the variable $u=1/r)$,
\bea A_1(u) &=& \frac{4 \pi}{3- \l}\frac{\rS}{\rDM} \frac{\rDM^3
\rhoDM}{M} \left( 1 - \rS u \right)
\int \frac{1}{\uDM} \frac{\left(u/\uDM \right)^{\l-3}}
{\left( 1-\rS u \right)^2}du, \label{A_1(u)}
\eea
and
\bea
B_1(u) &=& \frac{4\pi}{3-\lambda} \frac{1}{(1 - \rS u)^2}
\frac{\rS}{\rDM} \frac{\rDM^3 \rhoDM}{M} \left( \frac{u}{\uDM}
\right)^{\l - 2}\,, \label{B_1(u)} \eea
respectively.

\subsection{Perihelion precession}

In order to obtain the change in the perihelion precession due to
the presence of dark matter we need to evaluate the function $G(u)
= G_0(u) + G_1(u)$, where $G_0$ is the standard GR result, given in
(\ref{G_schw}), and $G_1$ is the first order perturbation generated
from the modification of the metric functions, which can be
written as
\be G_1 = \frac{B_1}{B_0^2}\left(u^2+\frac{1}{L^2}\right) -
\frac{E^2}{c^2L^2} \left( A_0 B_1 + A_1 B_0\right)\,. \label{G_1}
\ee
Plugging into this equation the expressions of the metric
functions (\ref{A_1(u)})-(\ref{B_1(u)}), we arrive at
\bea G_1 &=& \frac{4 \pi}{3- \l}\frac{\rS}{\rDM} \frac{\rDM^3
\rhoDM}{M} \left\{ \uDM^2 \left( \frac{u}{\uDM}\right)^{\l}
\right. \nn \\
&& \left. - \frac{E^2}{c^2L^2} \left[ \frac{1}{1-r_S u}
\left( \frac{u}{\uDM}\right)^{\l-2} -
\int \frac{1}{\uDM} \frac{\left(u/\uDM \right)^{\l-3}}
{\left( 1 - \rS u \right)^2}du \right] +
\frac{1}{L^2} \left( \frac{u}{\uDM}\right)^{\l-2} \right\}
\label{G_1_DM}\,,
\eea
and
\bea
F_1 &=& \frac{2 \pi}{3- \l}\frac{\rS}{\rDM} \frac{\rDM^3 \rhoDM}{M}
\left\{ \l \uDM \left( \frac{u}{\uDM}\right)^{\l-1}
-  \frac{E^2}{c^2L^2} r_S \left( \frac{u}{\uDM}\right)^{\l-2} \right.
\nn \\
&& \left. + \frac{E^2}{c^2L^2\uDM} \left[
-\l\left(1-\frac{c^2}{E^2}\right) +
\left(3-2\frac{c^2}{E^2}\right) \right]\left(
\frac{u}{\uDM}\right)^{\l-3} \right\}\,, \label{F_1_DM} \eea
respectively. In the small velocity limit, $ds \approx c dt$, so
that $E \approx c^2A_0\dot{t} \approx c$ (higher order would be
subleading in our expansion), and we can write
\be F_1 =  \frac{2 \pi}{3- \l}\frac{\rS}{\rDM} \frac{\rDM^3
\rhoDM}{M} \left[ \l \uDM \left( \frac{u}{\uDM}\right)^{\l-1} -
\frac{r_S}{L^2} \left( \frac{u}{\uDM}\right)^{\l-2} +
\frac{1}{L^2\uDM} \left( \frac{u}{\uDM}\right)^{\l-3} \right]\,,
\label{F_1_DM_corr} \ee

Using for $L$ and $u_0$ the values given
in Eqs.~(\ref{L})-(\ref{u0}) respectively, the perihelion precession angle can be written as
\bea
\s_1 &=& \frac{\pi}{3- \l}\frac{\rS}{\rDM} \frac{\rDM^3
\rhoDM}{M} \left\{ (\l^2-3\l-4) \left[
\frac{\rDM}{a(1-e^2)}\right]^{\l-2} + (\l-3) \frac{\rDM}{r_S}
\left[ \frac{\rDM}{a(1-e^2)}\right]^{\l-3}
\right\} \nn \\
& \approx & \pi  \frac{\rDM^3 \rhoDM}{M} \left[
\frac{\rDM}{a(1-e^2)}\right]^{\l-3}\,. \label{s_1_DM_corr} \eea

The observed value of the perihelion precession for the planet
mercury is $\d \phi_{O}=43.11 \pm 0.21$ arcsec per century
\cite{Sh}, while the predicted value from GR is $\d
\phi_{GR}=42.94$ arcsec per century. Assuming that the entire
discrepancy $\D \d \phi = 0.17 \pm 0.21$ is due to the presence of
dark matter, we obtain the following constraint for the mean density and the radius of the dark matter subhalo
\be \frac{\rDM^3 \rhoDM}{M_{\odot}} \left[
\frac{\rDM}{a(1-e^2)}\right]^{\l-3} \leq \frac{10^{-5}}{36^2
\pi}\frac{T_M}{T_E} \D \d \phi \,,\label{DM_per}
\ee
where $T_M$ and $T_E$ are, respectively,  the periods of revolution of Mercury and of the Earth.

\subsection{Light deflection}

To obtain the correction to the deflection of light given by the
presence of dark matter around the Sun, we must solve the general
equation (\ref{phot}). By defining $u = u_0 + u_1 + u_2$, where
$u_0$ is the straight line solution and $u_1$ is the GR result
(\ref{ulight}), we can write the equation as
\be \frac{d^2u_2}{d \phi^2} + u_2 = Q_1(u_0)
\,,\label{phot_DM_prov} \ee
with the functions $Q_1(u)$ and $P_1(u)$ obtained from the first
order corrections to the metric functions due to dark
matter\footnote{In the right hand side of Eq.~(\ref{phot_DM_prov})
we have dropped the term $Q_0(u_1)$, since it is of order
$(r_s/R)^3$, and can thus be neglected}. For the function $P_1$ we obtain
\be P_1 = \frac{B_1}{B_0^2}u^2 - \frac{E^2}{c^2L^2} \left( A_0 B_1
+ A_1 B_0\right) \,,\label{P_1} \ee
so that, by inserting the solutions (\ref{A_1(u)})-(\ref{B_1(u)}),
the functions $P_1$ and $Q_1$ are given by
\bea P_1 &=& \frac{4 \pi}{3- \l}\frac{\rS}{\rDM} \frac{\rDM^3
\rhoDM}{M} \Bigg\{ \uDM^2 \left( \frac{u}{\uDM}\right)^{\l}
 \nn \\
&& \left. - \frac{E^2}{c^2L^2} \left[ \frac{1}{1-r_S u} \left(
\frac{u}{\uDM}\right)^{\l-2} - \int \frac{1}{\uDM}
\frac{\left(\frac{u}{\uDM} \right)^{\l-3}}{\left( 1 - \rS u
\right)^2} du\right] \right\},
\label{P_1_DM} \\
Q_1 &=& \frac{2 \pi}{3- \l}\frac{\rS}{\rDM} \frac{\rDM^3
\rhoDM}{M} \left[ \l \uDM \left( \frac{u}{\uDM}\right)^{\l-1}
-  \frac{E^2}{c^2L^2} r_S \left( \frac{u}{\uDM}\right)^{\l-2} \right. \nn \\
&& \left. + (\l-3) \frac{E^2}{c^2L^2\uDM} \left(
\frac{u}{\uDM}\right)^{\l-3} \right]\,. \label{Q_1_DM} \eea
respectively.

Keeping only the leading order in the $r_S$ expansion, we can
again set $E^2/c^2 \approx 1$ and $L \approx R$. Then, inserting
the expression of $u_0$ in (\ref{Q_1_DM}), we can write Eq.~(\ref{phot_DM_prov}) as
\be \frac{d^2u_2}{d \phi^2} + u_2 =
\frac{2 \pi}{3- \l}\frac{\rS}{\rDM} \frac{\rDM^3 \rhoDM}{M} \left(
\frac{\rDM}{R}\right)^{\l-2} \left[ \l\cos(\phi)^{\l-1} -
(\l-3)\cos(\phi)^{\l-3} \right] \frac{1}{R}. \label{Phot_DM}
\ee
The solution of this equation is
\be u_2(\phi) = \frac{2 \pi}{3- \l}\frac{\rS}{\rDM} \frac{\rDM^3
\rhoDM}{M} \left( \frac{\rDM}{R}\right)^{\l-2} \left[
\frac{\cos(\phi)^{\l-1}}{\l-2} +2\sin(\phi) \int d\phi
\cos(\phi)^{\l-2} \right]. \label{phot_sol} \ee
The condition $u(\pi/2 + \e) = 0$ now yields
\be -\frac{\e}{R} + \frac{r_S}{R^2} + \frac{2 \pi}{3-
\l}\frac{\rS}{\rDM} \frac{\rDM^3 \rhoDM}{M} \frac{1}{R} \left(
\frac{\rDM}{R}\right)^{\l-2} \left[ \frac{5\l -
9}{(\l-1)(\l-2)}\e^{\l-1} + 2\sqrt{\pi} \frac{\Gamma(\l/2 -
1/2)}{\Gamma(\l/2)} \right] = 0, \label{eps_eq} \ee
where
\be
\G(z) = \int_0^{+\infty}\exp(-t)t^{z-1}
\label{Gamma}
\ee
is the Euler's $\G$ function \cite{Gradshteyn}.

Since we have assumed $\l > 2$, we can discard the term
$\e^{\l-1}$ in the previous equation, so that we are left with a
simple linear equation. Finally, the deflection angle can be
evaluated as
\be \d \phi = \frac{4GM}{c^2 R} \left[1 + \frac{8 \pi^{3/2}}{3-
\l}\frac{\rDM^3 \rhoDM}{M} \frac{\Gamma(\l/2 - 1/2)}{\Gamma(\l/2)}
\left( \frac{\rDM}{R}\right)^{\l-3} \right]\,. \ee
The best available data on light deflection by the Sun come from
long baseline radio interferometry \cite{all2}, which gives
$\delta \phi _{LD} = \delta \phi
_{LD}^{(GR)}\left(1+\Delta_{LD}\right) $, with $\Delta _{LD}\leq
0.0002 \pm 0.0008$ arcsec. Thus, assuming, as usual, that all the
discrepancy is due to the dark matter correction, and taking
$R=R_{\odot}$, we obtain the following constraint
\be \frac{8 \pi^{3/2}}{|3- \l|} \frac{\rDM^3 \rhoDM}{M_{\odot}}
\frac{\Gamma(\l/2 - 1/2)}{\Gamma(\l/2)} \left(
\frac{\rDM}{R_{\odot}}\right)^{\l-3} < \Delta_{LD} \,.
\label{DM_light_def} \ee

\subsection{Radar echo delay}

To obtain the expression of the radar echo delay in the presence of the dark matter sub halo, we have to evaluate the integral (\ref{gen_red}) with the
corrected metric functions. Since we can set
\be \sqrt{\frac{B_0+B_1}{A_0+A_1}} \simeq \sqrt{\frac{B_0}{A_0}} +
\frac{1}{2} \left(\frac{B_1}{B_0} -\frac{A_1}{A_0} \right) \,,\ee
by neglecting higher order terms we obtain
\be \frac{1}{2} \left(\frac{B_1}{B_0} - \frac{A_1}{A_0} \right)
\simeq \frac{2 \pi(\l -1)}{(\l-2)(\l-3)} \frac{\rS}{\rDM}
\frac{\rDM^3 \rhoDM}{M} \left( \frac{u}{\uDM}\right)^{\l-2}.
\label{red_corr_func} \ee
Therefore the correction to the standard GR result can be set as
\bea \d T_1 &=& \frac{2 \pi(\l -1)}{(\l-2)(3-\l)} \frac{\rS}{\rDM}
\frac{\rDM^3 \rhoDM}{M} \left( \frac{\rDM}{R}\right)^{\l-2}
\frac{1}{c} \int_{-l_1/R}^{l_2/R} dz
\left( 1+z^2 \right)^{1-\l/2} \nn \\
&&= \frac{2\pi(\l -1)}{(\l-2)(3-\l)} \frac{\rS}{\rDM} \frac{\rDM^3
\rhoDM}{M}
\left(\frac{\rDM}{R}\right)^{\l-2} \frac{1}{c}\times \nn \\
&& \times
\left[l_1~{}_2F_1\left(\frac{1}{2},\frac{\l}{2}-1,\frac{3}{2},-\left(
\frac{l_1}{R} \right)^2 \right) +
l_2~{}_2F_1\left(\frac{1}{2},\frac{\l}{2}-1,\frac{3}{2},-\left(
\frac{l_2}{R} \right)^2 \right) \right] \,,\label{red_DM} \eea
where ${}_2F_1(a,b,c,z)$ is the Hypergeometric function \cite{Gradshteyn}, defined as the solution of the
differential equation
\be
z(1-z)\frac{d^2w}{dz^2} + \left[c - (a+b+1)z \right]\frac{dw}{dz} - abw=0
\label{Hypdeq}
\ee

Observational constraints on the radar echo delay comes form the
frequency shift of radio photons to and from the Cassini
spacecraft \cite{Reasemberg}. We have $\D t_{RD} = \D
t_{RD}^{(GR)}(1+\D_{RD})$, with $\D_{RD} \simeq (1.1 \pm 1.2)
\times 10^{-5}$. Therefore, assuming as usual that the discrepancy
is completely due to the presence of dark matter, we obtain the constraint
\bea && \frac{2 \pi(\l -1)}{(\l-2)|3-\l|} \frac{\rDM^3
\rhoDM}{M_\odot} \left( \frac{\rDM}{R_{\odot}}\right)^{\l-2}
\left[\log\left(\frac{4l_1 l_2}{R^2}\right)\right]^{-1} \times \nn \\
&&\times \left[
\frac{l_1}{\rDM}~{}_2F_1\left(\frac{1}{2},\frac{\l}{2}
-1,\frac{3}{2},-\left( \frac{l_1}{R} \right)^2 \right) +
\frac{l_2}{\rDM}~{}_2F_1\left(\frac{1}{2},\frac{\l}{2}
-1,\frac{3}{2},-\left( \frac{l_2}{R} \right)^2 \right)
\right] < \D_{RD}. \nn \\
\label{Delta_RD_DM}
\eea

The next Section will be devoted to some comments on the above results, and we will draw our conclusions.

\begin{figure}[h]
\includegraphics[height=7.8cm]{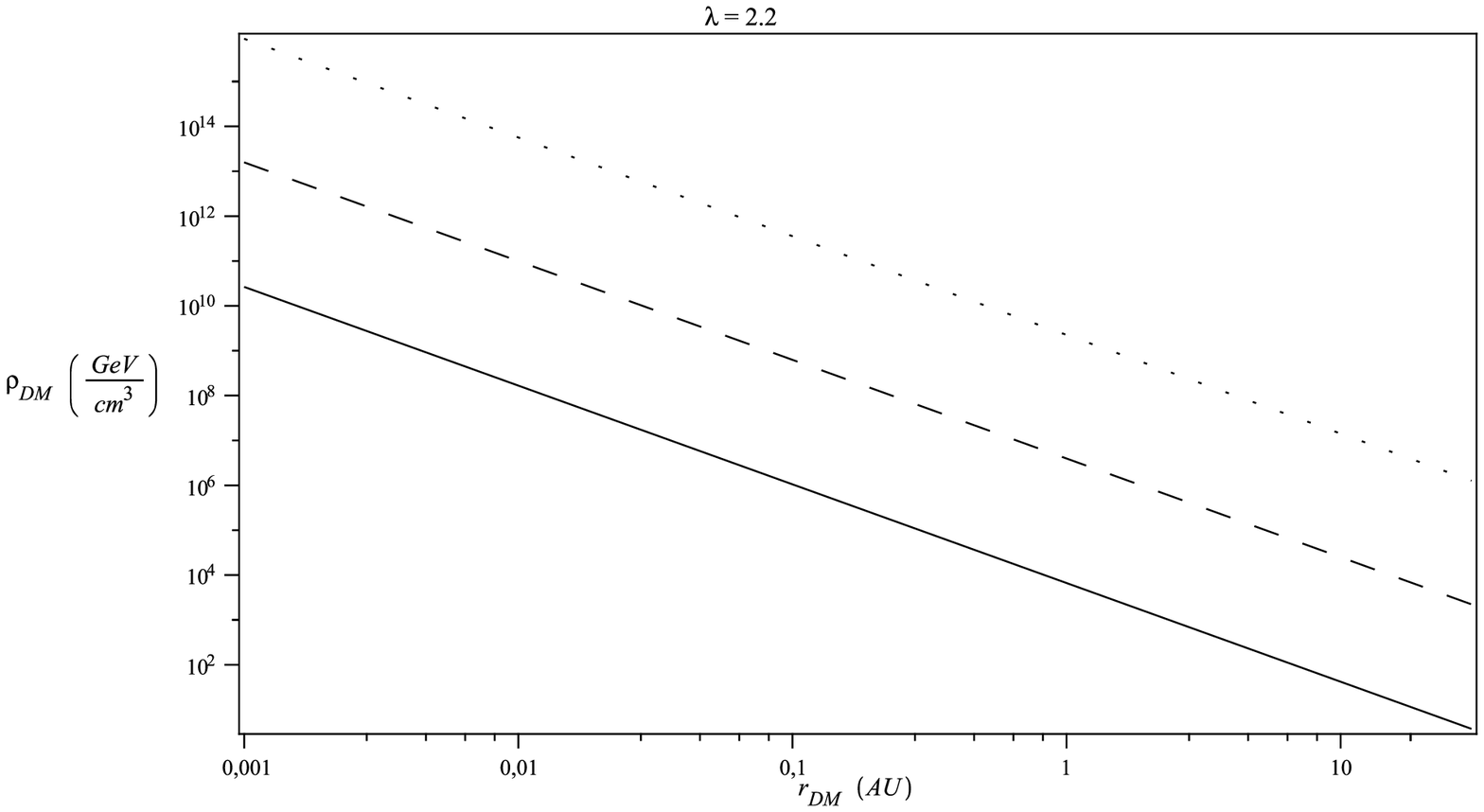}
\hspace*{0.2cm}
\includegraphics[height=7.8cm]{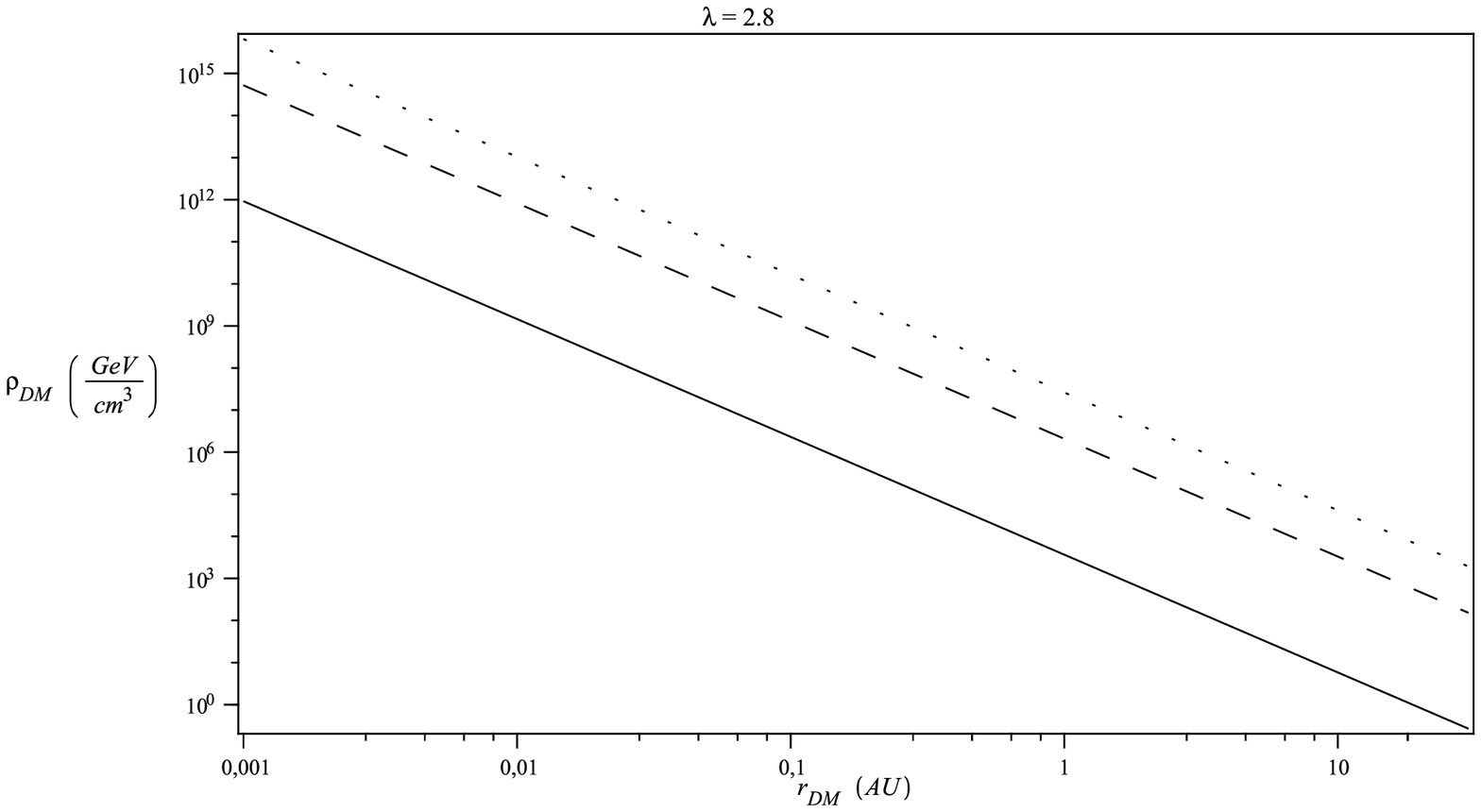}
\caption{Plots of the constraints on the parameters $\rDM$ versus
$\rhoDM$ of the dark matter distribution $\rho_1$ of Eq.
(\ref{DM_profile}) obtained from the perihelion precession (solid
line), the deflection of light (dashed line) and the radar echo
delay (dotted line). The values of $\l$ have been set to $\l=2.2$
for plot (a) and to $\l=2.8$ for plot (b). The radius is measured
in astronomical units, while the energy density is measured in ${\rm GeV/cm^3}$. (as a reference for
the reader, we remind that the mean value of the galactic dark matter halo is of order 1  ${\rm GeV/cm^3}$)}
\label{r_vs_rho}
\end{figure}
\begin{figure}[h]
\begin{center}
\includegraphics[height=7.8cm]{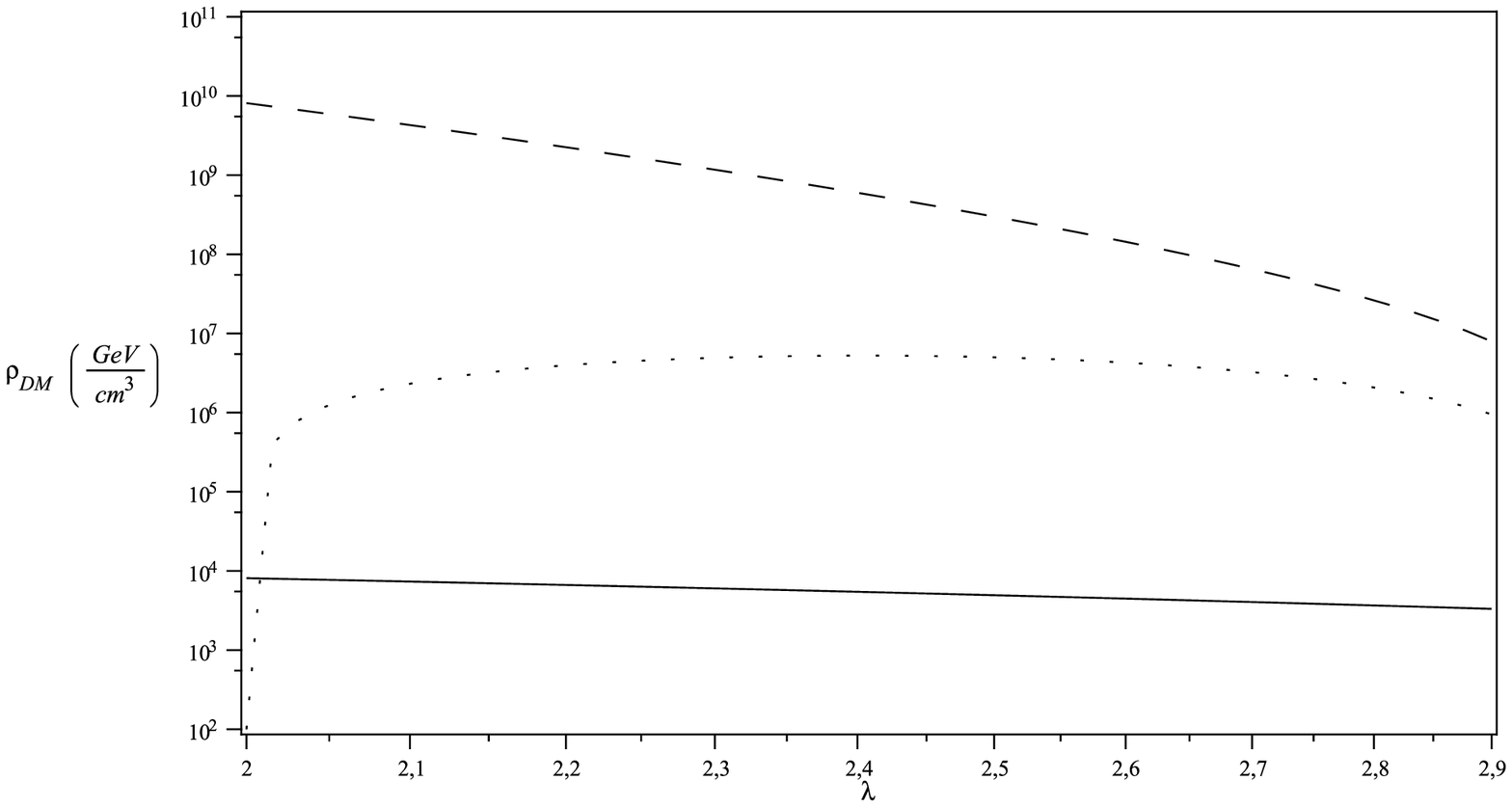}
\caption{Plots of the constraints on the parameters $\rhoDM$
versus $\l$ of the dark matter distribution $\rho_1$ of Eq.
(\ref{DM_profile}) obtained from the perihelion precession (solid
line), the deflection of light (dashed line) and the radar echo
delay (dotted line). The value of $\rDM$ have been set to $\rDM
\simeq 1 AU$. The energy density is measured in $GeV/cm^3$ (as a reference for
the reader, we remind that the mean value of the galactic dark matter halo is of order 1  ${\rm GeV/cm^3}$)}
\label{lambda_vs_rho}
\end{center}
\end{figure}

\section{Comments and conclusions}
\label{Comments}

In the previous Section we have evaluated the possible influence
that the presence of Sun-bound dark matter could have on the
classical tests of general relativity. The results we obtained are
summarized in Figs.~\ref{r_vs_rho} and \ref{lambda_vs_rho}.

As we can see, the most constraining test is always the study of
the perihelion precession. The results strongly depends on the
extension of the dark matter halo (Fig. \ref{r_vs_rho}). If the
distribution is extremely concentrated around the Sun, the effect
become more and more negligible. On the contrary, for halos
extending roughly to the orbit of the Earth, the constraint on the
mean dark density becomes comparable with the results obtained by
other authors \cite{Sereno:2006mw,Iorio:2006cn,Frere:2007pi}. For
halos as big as the planetary solar system, the constraint on the
mean dark matter energy becomes tighter. On the contrary, at least
for what concerns the perihelion precession, the $\l$-dependence
is quite mild (Fig.~\ref{lambda_vs_rho}). Notice that for very
large values of $\rDM$ the bounds obtained becomes comparable with
what is expected from the mean energy density distribution of the
galaxy \cite{Catena:2009mf}. This is a consequence of the choice
of our hypothesis for the energy density distribution
(\ref{DM_profile}), which do not constraint the total amount of
dark matter around the Sun. Nevertheless, we can argue that, if
some sort of overdensity can be localized around the Sun, a
distribution packed close to the Sun seems to be less in tension
with observations.

Some considerations on how the analysis we have performed can be
improved seems appropriate. Of course, assuming that the dark
matter distribution is static is certainly not true. Probably a
model of rotating dark matter could improve the constraints, but
the metric would be quite complicated. In fact, no exact non-empty
spherically symmetric rotating solution is known. As a future work, it could be possible to
perform a similar analysis, using the Kerr geometry as the
background expansion. For what concerns the profile energy
density, we believe that the expression we have used could capture
the essential features of a more realistic energy density
distribution, especially because we are mainly interested in the
dimension and the mean density of such a distribution. From this
point of view, we think that our results are robust. A different
but still interesting issue would be to take into account the
possibility of dark matter ``streams'' of extragalactic objects
recently captured by our galaxy, which could give some strong
local overdensities, and which could be characterized by a
definite velocity pattern. Maybe it could be possible to model
such an effect, for example via an Einstein cluster approach \cite{Eclus}. This
will also be the object of future investigations.

In conclusion, our analysis suggest that indirect gravitational
effect originated by dark matter can be interesting even at small
scales, and can be a useful cross-check for direct detection
experiments.

\section*{Acknowledgments}

We would like to thank Pasquale Dario Serpico for useful
discussion, and Maurizio Gasperini for helpful comments on the
manuscript. The work of TH was supported by an GRF grant of the
government of the Hong Kong SAR. FSNL acknowledges financial
support of the Funda\c{c}\~{a}o para a Ci\^{e}ncia e Tecnologia
through the grants CERN/FP/123615/2011 and CERN/FP/123618/2011.

\appendix

\section{Classical tests of General Relativity in an arbitrary spherically
symmetric static space-time} \label{general}

In this Appendix, we briefly review the formalism used in studying
the classical tests of general relativity. This formalism is
completely general, and can be used to treat any static
spherically symmetric metric \cite{Harko:2009qr,arXiv:0910.3800}.
The Appendix follows closely the line of the analogous Section of
\cite{arXiv:0910.3800}, and we include it for self-completeness
and self-consistency. The general metric of the form
\be\label{A1}
ds^2 = A(r) c^2dt^2 -B(r) dr^2 -r^2 d\Om^2 \,,
\ee
which will be used to obtain a general expressions for the
classical tests of general relativity. For the Schwarzschild metric the functions $A$ and $B$ are given by Eqs.~(\ref{schw_metric}).
%
%

\subsection{Perihelion precession}

The motion of a test particle in the gravitational field described by the
metric given by Eq. (\ref{A1}) can be derived from the
variational principle \be \delta \int
\sqrt{A(r)c^{2}\dot{t}^{2}-B(r)\dot{r}^{2}-r^{2}\left(
\dot{\theta}^{2}+\sin ^{2}\theta \dot{\phi}^{2}\right) } ds=0,
\label{var} \ee where the dot denotes $d/ds$. It is easy to see
that the orbit must be planar, and hence we can set $\theta =\pi
/2$ without any loss of generality. Therefore we will use $\phi $
as the angular coordinate. Since neither $t$ nor $\phi $ appear
explicitly in Eq. (\ref{var}), their conjugate momenta yield
constants of motion (related respectively with energy and angular
momentum conservation) \be A(r)c^{2}\dot{t}=E, \qquad
r^{2}\dot{\phi}=L. \label{consts} \ee

From (\ref{var}), using the constants of motion (\ref{consts}),
changing the variable $u = 1/r$ and expressing the affine
derivative as an angular derivative $d/ds=Lu^{2}d/d\phi$ we arrive
at the equation
\be \left( \frac{du}{d\phi }\right)^{2} + u^{2} =
\frac{B(u) - 1}{B(u)} u^{2} + \frac{E^{2}}{c^{2}L^{2}}
\frac{1}{A(u) B(u)} - \frac{1}{L^{2}}\frac{1}{B(u)}\equiv G(u).
\label{ueq_basic} \ee

By taking the derivative of the previous
equation with respect to $\phi $ we find \be \frac{d^{2}u}{d\phi
^{2}}+u=F(u), \label{inter2} \ee where \be
F(u)=\frac{1}{2}\frac{dG(u)}{du}. \ee

A circular orbit $u=u_{0}$ is given by the root of the equation
$u_{0} = F\left( u_{0}\right) $. Any deviation $\delta =u-u_{0}$
from a circular orbit must satisfy the equation \be
\frac{d^{2}\delta }{d\phi ^{2}} + \left[ 1-\left(
\frac{dF}{du}\right)_{u=u_{0}}\right] \delta =O \left( \delta
^{2}\right), \label{gen_dev} \ee which is obtained by substituting
$u=u_{0}+\delta $ in Eq. (\ref{inter2}). Therefore, to first order
in $\delta $, the trajectory is given by \be \delta =\delta _{0}
\cos \left( \sqrt{1-\left(\frac{dF}{du}\right)_{u=u_{0}} }\phi +
\beta \right), \label{gen_per} \ee where $\delta _{0}$ and $\beta
$ are constants of integration. The angles of the perihelia of the
orbit are the angles for which $r$ is minimum and hence $u$ or
$\delta $ is maximum. Therefore, the variation of the orbital
angle from one perihelion to the next is \be \phi
=\frac{2\pi}{\sqrt{1-\left( \frac{dF}{du}\right)_{u=u_{0}}}} =
\frac{ 2\pi }{1-\sigma }. \label{prec} \ee

The parameter $\sigma $ defined by the above equation represents the
perihelion advance, giving the rate of the
perihelion change. As the planet advances through $\phi $ radians in its
orbit, its perihelion advances through $\sigma \phi $ radians.
From Eq. (\ref{prec}), $\sigma $ is given by \be \sigma
=1-\sqrt{1-\left( \frac{dF}{du}\right) _{u=u_{0}}}, \ee or, for
small $\left( dF/du\right) _{u=u_{0}}$, by \be \sigma
=\frac{1}{2}\left( \frac{dF}{du}\right) _{u=u_{0}}. \ee

For a complete rotation we have $\phi \approx 2\pi (1+\sigma)$,
and the advance of the perihelion is $\delta \phi =\phi -2\pi
\approx 2\pi \sigma $. In order to be able to perform effective
calculations of the perihelion precession we need to know the
expression of $L$ as a function of the orbit parameters. If the planet is moving on a Keplerian ellipse with
semi-axis $a$, and eccentricity of the orbit is $e$, then \cite{Harko:2009qr,arXiv:0910.3800}
\be
L^{2}=\frac{GMa\left(1-e^{2}\right)}{c^2}. \label{L} \ee

As an example of the  application of the present formalism, and
for future references, we consider the precession of the
perihelion of a planet in the Schwarzschild geometry, where we
have \bea G(u) &=&
\frac{2GM}{c^{2}}u^{3}+\frac{1}{L^{2}}\left(\frac{E^{2}}{c^{2}}
-1\right) +\frac{2GM}{c^{2}L^{2}}u,
\label{G_schw} \\
F(u) &=&3 \frac{GM}{c^{2}}u^{2}+\frac{GM}{c^{2}L^{2}}.
\label{F_schw} \eea

The radius of the circular orbit $u_{0}$ is obtained as the
solution of the quadratic algebraic equation \be
u_{0}=3\frac{GM}{c^{2}}u_{0}^{2}+\frac{GM}{c^{2}L^{2}}, \ee with
the only physical solution given by \be u_{0}=\frac{1 \pm
\sqrt{1-12G^{2}M^{2}/c^{4}L^{2}}}{6GM/c^{2}}\approx \frac{GM
}{c^{2}L^{2}}. \label{u_0} \ee

Therefore \be \delta \phi =\pi \left( \frac{dF}{du}\right)
_{u=u_{0}}=\frac{6\pi GM}{ c^{2}a\left( 1-e^{2}\right) }, \ee
which is the standard general relativistic result \cite{LaLi}.

\subsection{Deflection of light}

In the absence of external forces a photon follows a null
geodesic, $ ds^{2}=0 $. The affine parameter along the photon's
path can be taken as an arbitrary quantity, and we denote again by
a dot the derivatives with respect to the arbitrary affine
parameter. There are two constants of motion, the energy $E$ and
the angular momentum $L$, given again by eqs. (\ref{consts}).

The equation of motion of the photon is \be
\dot{r}^{2}+\frac{1}{B(r)}r^{2}\dot{\phi}^{2}=\frac{A(r)}{B(r)}c^{2}\dot{t}
^{2}, \ee which, with the use of the constants of motion, the
change of variable $r=1/u$ and the use of the conservation
equations to eliminate the derivative with respect to the affine
parameter leads to \be \left( \frac{du}{d\phi }\right)^{2} +u^{2}
= \frac{B(u)-1}{B(u)} u^{2} + \frac{E^{2}}{c^2L^{2}} \frac{1}{A(u)
B(u)} \equiv P(u). \label{P_eq_basic} \ee By taking the derivative
of the previous equation with respect to $\phi $ we find \be
\frac{d^{2}u}{d\phi ^{2}} + u = Q(u), \label{phot} \ee where \be
Q(u)=\frac{1}{2}\frac{dP(u)}{du}. \ee

In the lowest approximation, in which the term of the right hand
side of the equation (\ref{phot}) is neglected, the solution is a
straight line, \be u = \frac{\cos \phi }{R}, \label{u0} \ee where
$R$ is the distance of the closest approach to the mass. In the
next approximation Eq. (\ref{u0}) is used on the right-hand side
of Eq. (\ref {phot}), to give a second order linear inhomogeneous
equation of the form \be \frac{d^{2}u}{d\phi ^{2}} + u =
Q\left(\frac{\cos \phi }{R}\right). \label{uQ} \ee with a general
solution given by $u=u\left( \phi \right) $. The light ray comes
in from infinity at the asymptotic angle $\phi = -\left( \pi
/2+\varepsilon \right)$ and goes out to infinity at an asymptotic
angle $ \phi =\pi /2+\varepsilon $. The angle $\varepsilon $ is
obtained as a solution of the equation $u\left(\pi /2+\varepsilon
\right) =0$, and the total deflection angle of the light ray is
$\delta =2\varepsilon $.

In the case of the Schwarzschild metric we have
\bea
P(u)=\left(2GM/c^{2}\right) u^{3}, \label{P_schw}
\eea
and
\bea
Q(u)=\left( 3GM/c^{2}\right) u^{2}, \label{Q_schw}
\eea
respectively. In the lowest approximation order from Eqs.
(\ref{u0}) and (\ref{uQ}) we obtain the second order linear
equation \be \frac{d^{2}u}{d\phi
^{2}}+u=\frac{3GM}{c^{2}R^{2}}\cos ^{2}\phi =\frac{3GM}{
2c^{2}R^{2}}\left( 1+\cos 2\phi \right), \ee with the general
solution given by \be u=\frac{\cos \phi
}{R}+\frac{3GM}{2c^{2}R^{2}}\left(1-\frac{1}{3}\cos 2\phi \right).
\label{ulight} \ee

By substituting $\phi =\pi /2 + \varepsilon $, $u=0$ into Eq.
(\ref{ulight}) we obtain the general relativistic result for the light deflection angle \cite{LaLi}
\be \delta =2\varepsilon = \frac{4GM}{c^{2}R}. \label{ld_Schw} \ee

\subsection{Radar echo delay}

The time travel of a light signal changes in the presence of a
gravitational field. The time difference can be cast as \be \delta
T = T-T_0 =
\frac{1}{c}\int_{-l_{1}}^{l_{2}}\left\{\sqrt{\frac{B(r)}{A(r)}}-1\right\}
dy. \label{gen_red} \ee where we can set $r = \sqrt{y^2 + R^2}$,
with $R$ being the distance of closest approach.

In the case of the Schwarzschild metric we have \be
\sqrt{\frac{B(r)}{A(r)}} = B(r) \simeq 1 + \frac{2GM}{c^2 r}. \ee
By assuming that the sources are far enough,
$R^2/l_{1}^{2}\ll 1$ and $R^2/l_{2}^{2}\ll 1$, and we have \cite{Sha}
\be \delta T
= \frac{2GM}{c^{3}} \int_{-l_{1}}^{l_{2}}
\frac{dy}{\sqrt{y^{2}+R^{2}}} = \frac{2GM}{c^{3}}\ln
\frac{4l_{1}l_{2}}{R^{2}}. \label{red_Schw} \ee

\end{document}